\documentclass[pra,reprint,aps,showpacs,merge,floatfix]{revtex4-1}
\usepackage{times,amsmath,graphicx,latexsym,dcolumn}

\begin{document}
\title{Vortex-loop calculation of the specific heat of superfluid $^4$He under pressure}

\author{Andrew Forrester}
\author{Gary A. Williams}
\affiliation{Department of Physics and Astronomy, University of California, Los Angeles, CA 90095}
\date{\today}  
\begin{abstract}
Vortex-loop renormalization is used to compute the specific heat of superfluid $^4$He near the lambda point at various pressures up to 26 bars.  The input parameters are the the pressure dependence of  T$_\lambda$ and the superfluid density, which determine the non-universal parameters of the vortex core energy and core size.  The results for the specific heat are found to be in good agreement with experimental data, matching the expected universal pressure dependence to within about 5$\%$.  The non-universal critical amplitude of the specific heat is found to be in reasonable agreement, a factor of four larger than the experiments.  We point out problems with recent Gross-Pitaevskii simulations that claimed the vortex-loop percolation temperature did not match the critical temperature of the superfluid phase transition.
\end{abstract}

\maketitle

\subsubsection{Introduction}
The idea that vortex loops could be the thermal excitations responsible for the superfluid $\lambda$-transition was first intuitively suggested by Onsager \cite{onsager}, in the same conference paper where he proposed the quantization of circulation in superfluids.  Much the same speculative idea was proposed by Feynman at the end of his 1955 review paper on superfluids \cite{feynman}, and both of them proposed that Landau's roton excitations were the smallest loops being excited, the ``ghosts of vanishing vortex rings".  The first actual calculations of a vortex-mediated superfluid transition were carried out by Kosterlitz and Thouless (KT) \cite{kt,kosterlitz} for two-dimensional (2D) superfluid films, with vortex-antivortex pairs the 2D equivalents of three-dimensional (3D) vortex loops.  The first 3D renormalization-group theory using circular vortex loops was proposed by one of us \cite{gw}.  This theory resulted in a 3D lambda-type phase transition where the superfluid density went to zero as a critical power law of the temperature difference from $T_{\lambda}$, and the specific heat peaked at the same temperature, though the critical exponents did not match known values.  This was quite different behavior from the 2D KT transition, where there is a sudden jump to zero of the superfluid density at $T_{KT}$, and no change in the specific heat at that point.

 The recursion relations resulting from the circular-loop theory were later verified by Shenoy \cite{shenoy} using a duality transformation of the Landau-Ginzburg-Wilson Hamiltonian, which diagonalizes exactly into vortex loop and spin-wave thermal excitations.   He was also able to show that the circular-loop ideas could be generalized to the actual random-walking loops, by employing a renormalized core size that was basically the extent of the fluctuations from the circular form \cite{chattopadhyay}.  This advance now brought the critical exponents of the loop theory into agreement with known values of the 3D O(2) universality class.

In this paper we calculate the specific heat from the vortex loops as a function of pressure, and find that the theory correctly predicts the universal scaling with pressure seen in the experiments.  The magnitude of the critical specific heat, though not universal, shows that the loops can easily account for nearly all of the entropy found very close to the $\lambda$-point.  We use this result to address a claim made recently \cite{cug} that the percolation temperature of the loops can be as much as 6\% lower than the critical temperature $T_c$.  If that were the case the simulations should have seen \textit{two} specific heat peaks, one at the percolation temperature and one at $T_c$.  Since they see only a single peak, the percolation point and $T_c$ must be identical, at least to within the precision of the simulation.

\subsubsection{Vortex-loop theory}
The loop theory starts from the hydrodynamic expression for the energy of single circular loops of diameter $a$, divided by the thermal energy $k_B T$, 
\begin{equation}
{{{U_0}} \mathord{\left/
 {\vphantom {{{U_0}} {{k_B}T}}} \right.
 \kern-\nulldelimiterspace} {{k_B}T}} = {\pi ^2}{K_0}\left( {a/a_0} \right)\left( {\ln (a/{a_c}) + C} \right)\quad,
\end{equation}
where 
\begin{equation}
{K_0} = \frac{{{\hbar ^2}\rho _s^0{a_0}}}{{{m^2}{k_B}T}}\quad,
\end{equation}
with $\rho_s^0$ the unrenormalized (``bare") superfluid density, $m$ the atomic mass, and $a_0$ the bare core diameter.  The parameter C gives the energy of the vortex core, while the log term is from the kinetic energy of the flowing helium.  $a_c$ is an effective core size, and for the bare case is proportional to $a_0$.

For the case of many interacting loops, the renormalized loop theory builds on two fundamental ideas used initially by Kosterlitz and Thouless \cite{kt} in 2D.  Vortex loops are dipolar objects, and orient in an applied flow such that their backflow opposes the applied flow.  The superfluid density is the net mass per unit volume flowing in response to the applied flow, and hence it is reduced by the polarizablility of the dipoles, which varies as $a^4$.  Linear response theory leads to an integral equation for the scale-dependent renormalized superfluid density $\rho_s^r (a)$,
\begin{equation}
\frac{1}{{{K_r}(a)}} = \frac{1}{{{K_o}}} + \frac{{4{\pi ^3}}}{{3a_o^7}}\int_{{a_o}}^a {{a^4}\,} {e^{ - U(a)/{k_B}T}}\,{\kern 1pt} {a^2}{\kern 1pt} da\quad,
\end{equation}
where $K_r /K_0 = \rho_s^r/\rho_s^0$.
$U(a)$ is the renormalized loop energy, where the energy of a large loop is screened by the smaller loops  around it, and following the ``dielectric" screening ideas of Kosterlitz and Thouless gives
\begin{equation}
\begin{gathered}
  \frac{{U(a)}}{{{k_B}T}} = \int_{{a_o}}^a {\left( {\frac{1}{{{K_o}/{K_r}(a)}}} \right)} \frac{{\partial ({U_o})}}{{\partial a}}da + {\pi ^2}{K_o}C \hfill \\
= {\pi ^2}\int_{{a_o}}^a {{K_r}(a)\left( {\ln \left( {\frac{a}{{{a_c}}}} \right) + 1} \right)\;} \left( {\frac{{da}}{{{a_o}}}} \right) + {\pi ^2}{K_o}C \hfill \\\;.
\end{gathered} 
\end{equation}

These two coupled integral equations can be iterated to macroscopic scales, but it is more convenient to convert them to differential recursion relations using a length scale $\ell = \ln (a/a_0)$ and fugacity $y = (a/a_0)^6 \exp(-U(a)/k_B T)$: 
\begin{equation}
  \frac{{\partial K}}{{\partial \ell }} = K - \frac{{4{\pi ^3}}}{3}{K^2}y
\end{equation}
\begin{equation}
 \frac{{\partial y}}{{\partial \ell }} = \left( {6 - {\pi ^2}K\left( {\ln (a/{a_c}) + 1} \right)} \right)y
\end{equation}
\begin{equation}
 {K_r} = K{e^{ - \ell }}\quad.
\end{equation}
These can be iterated starting from the $\ell = 0$ values $K_0$ and
$y_0 = \exp ( - {\pi ^2}{K_0}C)$, and the macroscopic superfluid density is found from $K_r$ as 
$\ell  \to \infty$. The correlation length of the theory is given by 
\begin{equation}
\xi  = \frac{{{a_0}}}{{{K_r}}} = \frac{{{m^2}{k_B}T}}{{{\hbar ^2}{\rho _s}}}\quad ,
\end{equation}
the known value \cite{ahlerscor}.  
The effective core parameter $a_c$ is found using a free-energy minimization of the fluctuations about the minimum-energy circular configuration \cite{chattopadhyay}, and is given by  
\begin{equation}
{{{a_c}} \mathord{\left/
 {\vphantom {{{a_c}} a}} \right.
 \kern-\nulldelimiterspace} a} = {K^\theta }\quad,
\end{equation}
with the exponent $\theta = d/((d+2)(d-2))$ in $d$ dimensions \cite{gwtheta} near $d = 3$.
The fixed points of the recursion equations using $\theta = 0.6$ in $d = 3$ are $
{K^ * } = 0.387508\ldots$ and $
{y^ * } = 0.062421\ldots$ .  The temperature is varied by changing $K_0$, and the critical value $K_{0c}$ is determined when $K$ and $y$ scale to their fixed-point values at large $\ell$;  $y$ blows up and $K$ scales rapidly to zero at values of $K_0 < K_{0c}$.  The value of $K_{0c}$ is completely dependent on the value of C, through the initial value of $y_0$.

Expanding the recursion relations about the fixed points gives an analytic expression for the correlation-length exponent \cite{shenoy},
\begin{equation}
\frac{1}{\nu } = \frac{1}{2}\left( {\sqrt {1 + 24\left( {1 - \frac{{{\pi ^2}\theta {K^ * }}}{6}} \right)}  - 1} \right)\quad ,
\end{equation}
and putting in the $d=3$ values of ${K^ * }$ and $\theta$ gives $\nu = 0.67168835\ldots$ .  This is in complete agreement with the best high-temperature expansion \cite{vicari} value (0.6717(1)), the best Monte Carlo simulation \cite{machta} value  (0.6717(3)), and is within the bounds of the recent conformal bootstrap calculations \cite{conformal}.  We note that Eq.\,(7) guarantees that the loop theory exactly satisfies the Josephson hyperscaling relation \cite{josephson} for the critical exponent of the specific heat, $\alpha = 2-d\nu$, giving $\alpha = -0.015065\ldots$ for $d = 3$.

The critical exponent $\eta$ in the loop theory is effectively $\eta$ = 0 \cite{shenoy}, due to the neglect of small logarithmic terms in the duality transformation.  It is known to be small in 3D;  the most accurate simulation \cite{vicari} gives $\eta = 0.0381(2)$. In the loop theory the Hausdorff fractal dimension of the wandering loops \cite{gwD} is given by $D_H = 1/(1-\theta) = 2.50$, in exact agreement with the scaling theory \cite{chalker} prediction  $D_H = (5-\eta)/2$ with $\eta = 0$.

 \begin{figure}[t]
\begin{center}
\includegraphics[width=1.00\linewidth]{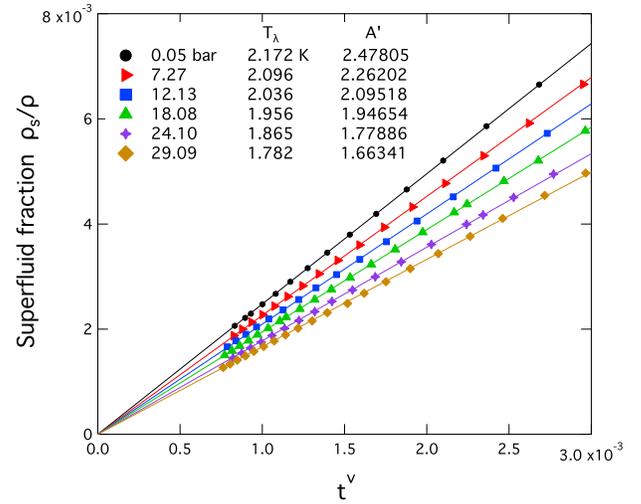}
\end{center}
\caption{(Color online) Fits to the data of Ref.\,\cite{greywall}, showing the values used for $T_{\lambda}$ and the resulting fit values of the amplitude $A'$}
\label{fig1}
\end{figure}

\subsubsection{Specific heat calculation}
To compare with experiments in superfluid $^4$He, it is necessary to determine the non-universal parameters of the core diameter $a_0$ and the core energy constant $C$ in the critical region near $T_\lambda$.  We use the experimental values of the critical amplitude of the superfluid density with pressure, and the pressure dependence of $T_\lambda$.  For the superfluid density we postulate that the starting value $\rho_s^0$ of the superfluid density at $\ell = 0$ is equal to the total density 
$\rho$.  It is well known that phonons contribute a negligible amount to the thermodynamics at the lambda point, particularly at higher pressures where the sound velocity increases, and in any event there are already larger noncritical contributions to the background specific heat from other sources, such as chemical-bond effects \cite{granato} that would be present even in the normal liquid.  We also do not include any ``roton" contributions to $\rho_s^0$, as we identify these excitations as the smallest vortex loops in the theory, as will be apparent shortly.  The recursion relations are iterated using standard fourth-order Runge-Kutta techniques to values of $\ell$ between 10 and 100, depending on the closeness to $T_{\lambda}$.

Figure 1 shows fits to the superfluid density measurements of Ref.\,\cite{greywall}.  We fit to the amplitude $A'$ in the form
\begin{equation}
\frac{{{\rho _s}}}{\rho } = A'{t^\nu } = \frac{{{K_r}}}{{{K_0}}}\quad,
\end{equation}
where $t = 1 - T/T_{\lambda}$, and in 3D the critical superfluid exponent is the same as the correlation-length exponent, from Eqs.\,(8) and (10).  We assume that $\nu = 0.67168835$ is universal and independent of pressure, and only allow $A'$ to vary, with results shown in the table in Fig.\,1.  This is different from the assumption in Ref.\,\cite{greywall}, who allowed also $\nu$ to vary, finding values close to 0.6717 but varying slightly with pressure.  Our values for $A'$ turn out to be consistently about 2-3\% higher than those of Ref.\,\cite{greywall}.
\begin{figure}[t]
\begin{center}
\includegraphics[width=1.00\linewidth]{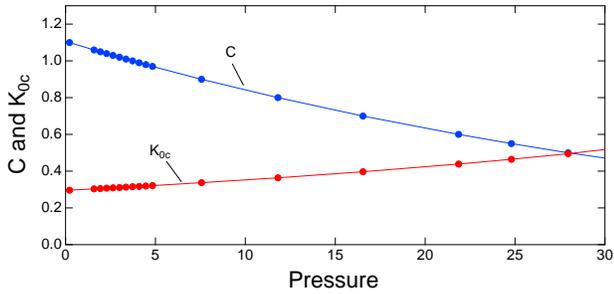} 
\end{center}
\caption{(Color online) The parameters $C$ and $K_{0c}$ versus pressure.  Additional data at low pressures  from Ref.\,\cite{greywall} were not shown in Fig.\,1.  The solid lines are polynomial fits to guide the eye.}
\label{fig2}
\end{figure}

Once $A'$ is known in the critical region, the corresponding values of $C$ can be found from iterating the recursion relations.  At a given pressure trial values of C and $K_0$ are inserted in $y_0$, and $K_r$ is calculated at long length scales from the recursion.  $K_0$ is then incremented to sweep the temperature, finding the critical value $K_{0c}$ (where $K_r$ goes to zero) and the superfluid fraction $K_r / K_0$ as a function of $T / T_{\lambda} =  K_{0c} / K_0$.  The resulting superfluid amplitude is then compared with $A'$, and if it does not match, the value of C is incremented, and the process is repeated until a match is found.  Figure 2 shows the resulting values of $C$ and $K_{0c}$ as a function of pressure.  The core energy in the critical region from Eq.\,(1), $U_0/ k_B = \pi^2 K_{0c} C\,T_{\lambda}$, is plotted in Fig.\,3 as a function of pressure, and also shown are the roton energy gaps at $T = T_{\lambda}$ from precision neutron scattering measurements under pressure \cite{gibbs}.  The agreement is seen to be quite good, further justifying our claim that the smallest loops can be treated as the roton excitations seen in experiments.
\begin{figure}[t]
\begin{center}
\includegraphics[width=1.00\linewidth]{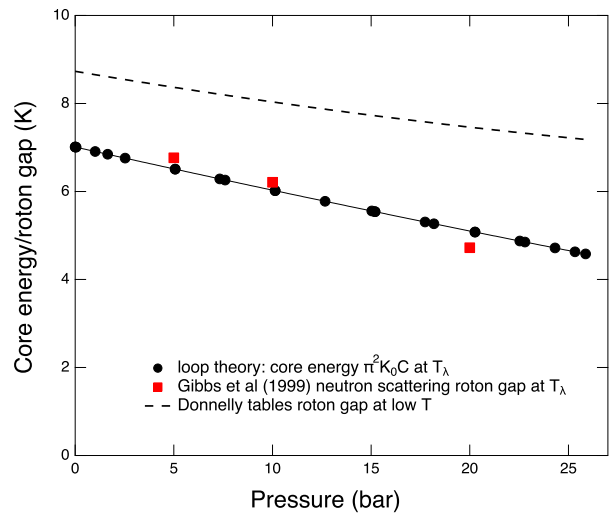}
\end{center}
\caption{Vortex core energy at $T_{\lambda}$, compared with the neutron scattering roton gap from Ref.\,\cite{gibbs}.  The solid fit line is a guide to the eye, while the dashed curve shows the roton gap energy at low temperature from Ref.\,\cite{donnellytables}.}
\label{fig3}
\end{figure} 

The value of $a_0(p)$ near the lambda point can now be found from $T_{\lambda}(p)$,
\begin{equation}
{a_0}(p) = \frac{{{m^2}{k_B}{T_\lambda }(p)}}{{{\hbar ^2}\rho (p)}}{K_{0c}}(p)\quad.
\end{equation}
At p = 0 bar this gives a value of $a_0 (0)$ = 2.41 {\AA}, in reasonable agreement with the value of 1.6 {\AA} deduced from vortex-ring measurements \cite{glaberson} at low temperatures (0.2 K).  The bare core diameter remains microscopic in the loop theory at the lambda point; it is the effective core size $a_c$ which becomes divergent there.  The pressure dependence of $a_0$ is shown in Fig.\,4, where it increases with pressure, though not quite as rapidly as found in low-temperature vortex-ring measurements.
\begin{figure}[t]
\begin{center}
\includegraphics[width=1.00\linewidth]{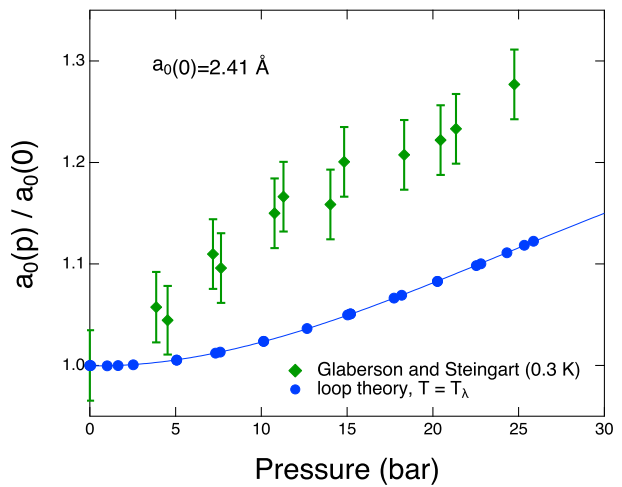}
\end{center}
\caption{(Color online) Vortex core diameter versus pressure.  The solid line is a fit to guide the eye.  The low-temperature measurements are from Ref.\,\cite{glaberson}.}
\label{fig4}
\end{figure} 

The loop recursion relation for the free energy was derived by Shenoy \cite{shenoy} from the duality transformation:
\begin{equation}
\frac{{\partial f}}{{\partial \ell }} =  - \pi \exp ( - 3\ell )y
\end{equation}
where $f = (F/k_BT)(a_0^3/V)$ with $F$ the free energy in the volume $V$, normalized by $a_0^3$. 
At constant pressure, the total heat capacity is 
\begin{equation}
{C_p} =  - T{\left( {\frac{{{\partial ^2}G}}{{\partial {T^2}}}} \right)_p} =  - T{\left( {\frac{{{\partial ^2}F}}{{\partial {T^2}}}} \right)_p} - T\,p{\left( {\frac{{{\partial ^2}V}}{{\partial {T^2}}}} \right)_p}\;,
\end{equation}
where $G = F + pV$ is the Gibbs free energy.  
The temperature derivatives of $V$ and $a_0$ can be rewritten in terms of derivatives of $\rho$, and we convert to the molar specific heat using $c_p = (V_m/V)C_p$, where $V_m = N_A m/\rho$ is the molar volume, with $N_A$ Avogadro's constant.  With the gas constant $R = k_B N_A$, the specific heat becomes (neglecting very small terms) 
\begin{widetext}
\begin{equation}
{c_{p,loop}} = \frac{{m\,R}}{{\rho \,a_0^3}}\left[ { - K_0^2{{\left( {\frac{{{\partial ^2}f}}{{\partial K_0^2}}} \right)}_p} + {K_0}{{\left( {\frac{{\partial f}}{{\partial {K_0}}}} \right)}_p}\left( {\frac{{4T}}{\rho }} \right){{\left( {\frac{{\partial \rho }}{{\partial T}}} \right)}_p} - f\left\{ {\left( {\frac{{2{T^2}}}{{{\rho ^2}}}} \right){{\left( {\frac{{{\partial ^2}\rho }}{{\partial {T^2}}}} \right)}_p} + \frac{{8T}}{\rho }{{\left( {\frac{{\partial \rho }}{{\partial T}}} \right)}_p}} \right\}} \right]\quad .
\end{equation}
\end{widetext}

By iterating Eqns.\,(5-7) and (13,15) for seven different values of pressure between 0.05 and 25.87 bars, we are able to get the calculated loop specific heat $c_{p,loop}$ versus temperature and pressure.  To analyze the temperature behavior we employ a fitting function
\begin{equation}
c_{p,fit} = B' t^{-\alpha} + D'\quad ,
\end{equation}
using the value of $\alpha$ given above in accordance with Josephson hyperscaling.  Fits to the calculated results in the critical regime at $t < 4\times 10^{-3}$ give values of the amplitude $B'_{loop}$ and the offset $D'_{loop}$, shown in Table I for the different pressures.  The experimental data shown in Fig.\,5 is also known to follow a similar form \cite{ahlers}, and we have fit the data to this form (again for $t < 4\times 10^{-3}$ and using our value $\alpha = -0.015065$), with the results for $B'_{exp}$ and $D'_{exp}$ shown in Table I.  
\begin{table}
\caption{\label{tab:example}Specific heat fit parameters, in units J/(mole K).}
\begin{ruledtabular}
\begin{tabular}{lllllcc}
Pressure (bar) & $B'_{loop}$ & $D'_{loop}$ & $B'_{exp}$ & $D'_{exp}$ & $B'_{exp}/B'_{loop}$ & $D''$\\
0.05 & -1397  & 1372 & -388.0 & 400.1 & 0.2777 & 19.14\\
1.65 & -1406  & 1381 & -382.1 & 394.0 & 0.2717 & 18.68\\
7.33 & -1401  & 1377 & -355.0 & 365.8 & 0.2534 & 16.76\\
15.03 & -1328  & 1308 & -337.5 & 346.6 & 0.2542 & 13.99\\
18.18 & -1286  & 1269 & -322.7 & 331.9 & 0.2510 & 13.48\\
22.53 & -1223  & 1210 & -299.6 & 309.5 & 0.2449 & 13.16\\
25.87 & -1173  & 1164 & -306.5 & 315.8 & 0.2613 & 11.76\
\end{tabular}
\end{ruledtabular}
\end{table}

\begin{figure}[t]
\begin{center}
\includegraphics[width=1.00\linewidth]{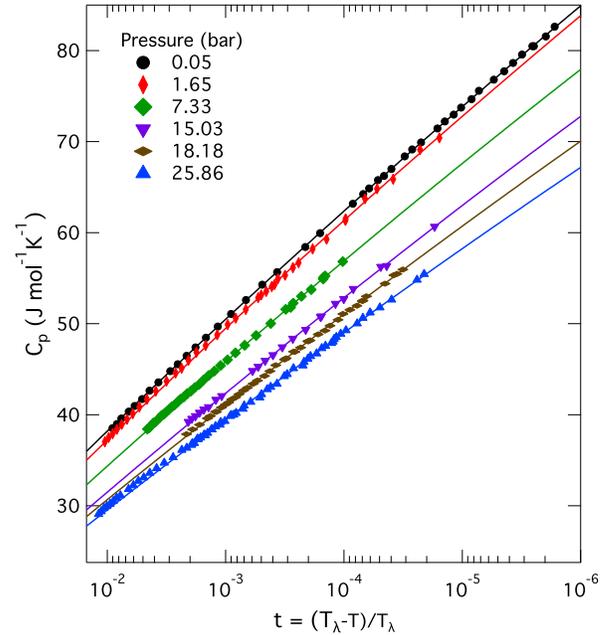}
\end{center}
\caption{Specific heat data from Ref.\,\cite{lipa} (0.05 bar) and all other data points from Ref.\,\cite{ahlers}, compared with the adjusted loop theory $c_{p,adj}$ (solid lines).}
\label{fig3}
\end{figure} 

Since the amplitude $B'_{loop}$ is a non-universal quantity that depends on other non-universal values at the bare scale $a_0$, we would not expect $B'_{loop}$  and $B'_{exp}$ to be the same, and indeed their ratio shown in Table I averages to a value of nearly 0.26.  What we do expect, however, is the pressure dependence of both amplitudes to match, due to universality arguments given by Rudnick and Jasnow \cite{rudnick} and Ferer \cite{ferer}.  Their ratio is nearly independent of pressure, within fluctuations of at most 5\% that are likely due to the uncertainties in the experimental data.   To compare more carefully with the data we plot as the solid lines in Fig.\,5 the theoretical specific heat adjusted by the amplitude ratios,
\begin{equation}
{c_{p,adj}} = ({B'_{\exp }}/{B'_{loop}})\,{c_{p,\,loop}} + D''\quad,
\end{equation}
where $D'' = D'_{exp} - (B'_{exp}/B'_{loop})D'_{loop}$ is the non-critical background probably coming from the chemical-bond effects \cite{granato}, and any non-loop excitations such as phonons.  The adjusted loop theory curves are seen to completely match the experimental data for $t < 4\times 10^{-3}$.  We note that this type of scaling adjustment of the amplitudes is also used in perturbative renormalization group calculations of the lambda specific heat \cite{dohm}, where again there is uncertainty in the precise values of the parameters at the microscopic bare length scale. 

A further test of universality is the parameter $X$ which was proposed by Ferer to be a universal constant independent of pressure \cite{ferer},
\begin{equation}
X = \frac{{\left| {B'} \right|}}{{R{V_m}}}{\left( {\frac{{{T_\lambda }{V_m}}}{{A'}}} \right)^3}\quad.
\end{equation}
In Table II we compare values of $X$ from the above values of $B'$ and $A'$ from the loop theory and the experiments, and with Ferer's values using the experimental data fit to 
$\alpha = -0.02$ and $\nu = 0.669$.  It is seen that the loop theory values are almost completely independent of pressure, while the experimental values are nearly constant, but fluctuate on the order of 5\%.  The loop values are larger by a factor of about 4, since $B'_{loop}/B'_{exp}\approx 4$. 
\begin{table}[h]
\caption{\label{tab:example}Universal parameter X, in units $10^{-4}$R(mole/cm$^3$)(K cm$^3$/mole)$^3$.  $X_{exp}$ values from our fits to Refs.\,\cite{ahlers} and \cite{greywall};  $X_{Ferer}$ values from Ref.\,\cite{ferer}.}
\begin{ruledtabular}
\begin{tabular}{llll}
Pressure (bar) & $X_{loop}$ & $X_{exp}$ & $X_{Ferer}$\\
0.05 & 8.68  & 2.41 & 2.06\\
1.65 & 8.68  & 2.36 & 2.02\\
7.33 & 8.68  & 2.20 & 1.99\\
15.03 & 8.67  & 2.21 & 2.02\\
18.18 & 8.67  & 2.18 & 1.99\\
22.53 & 8.66  & 2.12 & 1.96\\
25.87 & 8.66  & 2.26 & 2.19\
\end{tabular}
\end{ruledtabular}
\end{table}

\subsubsection{Conclusions}
These results show that the vortex-loop theory provides a complete description of the rapid rise of the specific heat of superfluid helium very close to $T_{\lambda}$.  The theory is entirely consistent with universality, and the critical exponents agree with the best simulation values.  Even the non-universal amplitude of the loop specific heat is the correct order of magnitude, only a factor of about four larger than the experimental value.  This difference likely stems from our use of the hydrodynamic form of Eq.\,(1) for the energy of the smallest loops at the bare scale, since with $a_0$ only a few {\AA}, quantum methods might give a better description \cite{roton}.  The amplitude of Eq.\,(15) varies rapidly with $a_0$, so only a small change in $a_0$ would be needed to better match with the experiments.

In this work the superfluid transition temperature $T_{\lambda}$ is exactly coincident with the temperature at which the loops percolate, e.g. for an infinite system the temperature where infinite diameter loops are just nucleated, made possible by the screening of the largest loops by all the loops of smaller size.  In a recent paper, Kobayashi and Cugliandolo \cite{cug} have claimed from Gross-Pitaevskii (GP) simulations that these two temperatures are different, with the loop percolation temperature falling below the superfluid transition temperature by as much as 6\%.  In coming to this conclusion they needed to study the loop distribution function in their simulations, the number of loops with a given diameter.  The problem with GP vortices is that the core size is relatively very large, spread out over several lattice points.  Near the transition the vortex density becomes very high, and it becomes very difficult to follow the trajectory of a loop when it comes close to another loop.  To resolve which loop is which they employed two different guesses in this situation: a ``stochastic" scenario in which the loop directions near a crossing are randomly chosen (which gave the percolation point 2\% below the superfluid $T_{c}$), and a second scenario where the configuration resulting in the longest loop is chosen  (percolation point 6\% below $T_{c}$).  The fact that the two different scenarios give quite different percolation points is already evidence that something is amiss with the procedures for determining the loop distributions.

We disagree with the conclusion that the two temperatures are different, and use their simulation of the GP specific heat and superfluid density as evidence this is not the case.  Their specific heat data (Figs.\,3b and 4c of Ref.\,\cite{cug}) shows only a single peak that is coincident with their superfluid transition.  If the loop percolation temperature was as much as 6\% smaller than $T_{c}$, our calculation shows that there should have been a large specific heat peak right at that point, and yet nothing at all is visible there in the data.  There is only the sharp peak right at $T_{c}$, which we and others \cite{sudbo} would argue is actually also the percolation point.  It is also impossible that the superfluid density would be unaffected by the loop percolation, and yet it shows no change at all at the purported lower percolation temperatures (Figs.\,6 and 7 of Ref.\,\cite{cug}).  In the loop theory the specific heat and superfluid density are only driven by the very longest loops, and are completely independent of the reconnection guesses discussed above.  It is only in computing aspects of the the loop distribution function that the inablility to accurately trace the loops will become troublesome, and we postulate this is the source of the claim in Ref.\,\cite{cug} that the loop percolation does not coincide with the superfluid $T_c$ \cite{note}.  We point out that recently a new procedure for accurately tracking the cores of GP vortices has been developed \cite{salman,giorgio}.  This procedure is able to completely track a reconnection event between two vortex lines that cross, following the details of the motion even at the point of closest approach.  It would be very interesting if this procedure could be applied to tracing the vortex loops in simulations like those of Ref.\,\cite{cug} to determine whether the loops do in fact percolate at the superfluid transition.

\begin{acknowledgments}
This work was supported in part by a grant from the Julian Schwinger Foundation.  We thank J. Lipa for sending us his experimental data, and acknowledge useful discussions with S. Putterman, J. Rudnick, and J. M. Kosterlitz.
\end{acknowledgments}


\begin{thebibliography}{32}
\bibitem [{gwt()}]{gwtheta}%
  \BibitemOpen
  \href@noop {} {\bibinfo  {journal} {In $d$ dimensions it is necessary to
  modify the result of Ref.\,\cite{chattopadhyay}, $\theta = d/(d+2)$, to
  $\theta = d/((d+2)(d-2))$. This comes about because
  Ref.\,\cite{chattopadhyay} takes the potential of a vortex segment to fall
  off as $1/R$, while in $d$ dimensions it will actually fall off as
  $1/R^{(d-2)}$. The modified value is only valid near $d = 3$, and still gives
  $\theta = 0.6$ in $d =3$}\ }\BibitemShut {NoStop}%
\bibitem [{not()}]{note}%
  \BibitemOpen
  \href@noop {} {\bibinfo  {journal} {As noted in Ref.\,\cite{cug} there have
  been previous claims from XY model simulations \cite{XY} that the vortex
  percolation is not coincident with the superfluid transition. In the XY model
  the vortex cores are better defined, and the claimed percolation temperature
  was found to be only about 0.3\% below the superfluid $T_c$ for the
  "stochastic" guess when two lines come close. As noted in the main text, it
  is unlikely that these results can be trusted without employing more accurate
  means of tracing the loops}\ }\BibitemShut {NoStop}%
\end{thebibliography}
\end{document}